\documentclass[aps,prb,twocolumn,amsmath,amssymb,superscriptaddress,floatfix]{revtex4}
\usepackage{graphicx}
\usepackage{graphics}
\usepackage{bm}
\usepackage[usenames]{color}
\usepackage{colordvi}
\usepackage{units}
\usepackage{bbm}
\usepackage{booktabs}
\usepackage{array}
\usepackage{placeins}
\usepackage{epsfig}
\usepackage{lscape}
\usepackage{changes}
\usepackage{braket}
\usepackage{tabularx}
\newcolumntype{L}[1]{>{\raggedright\arraybackslash}p{#1}} 
\newcolumntype{C}[1]{>{\centering\arraybackslash}p{#1}} 
\newcolumntype{R}[1]{>{\raggedleft\arraybackslash}p{#1}} 
\usepackage[colorlinks=true,linkcolor=blue,citecolor=blue,urlcolor=blue]{hyperref}
\newcommand{\be}{\begin{equation}}
\newcommand{\ee}{\end{equation}}
\newcommand{\beqn}{\begin{eqnarray}}
\newcommand{\eeqn}{\end{eqnarray}}

\definecolor{mymagenta}{rgb}{1.0,0.0,1.0}
\definecolor{mycyan}{rgb}{0.0,1.0,1.0}
\definecolor{myyellow}{rgb}{1.0,1.0,0.0}
\definecolor{myorange}{rgb}{1.0,0.27,0.0}

\definecolor{dark-gray}{HTML}{a0a0a0}
\definecolor{dark-red}{HTML}{8b0000}
\definecolor{dark-green}{HTML}{006400}
\definecolor{dark-blue}{HTML}{00008b}
\definecolor{gold}{rgb}{1.0,0.84,0.0}
\definecolor{dark-turquoise}{HTML}{00ced1}

\bibstyle{apsrev.bib}

\begin{document}

\title{Reentrant Random Quantum Ising Antiferromagnet}
\author{P\'eter Lajk\'o}
\email{peter.lajko@ku.edu.kw}
\affiliation{Department of Physics, Kuwait University, P.O. Box 5969, Safat 13060, Kuwait}
\author{Jean-Christian Angl\`es d'Auriac}
\email{dauriac@neel.cnrs.fr}
\affiliation{Institut N\'eel-MCBT CNRS, B. P. 166, F-38042 Grenoble, France} 
\author{Heiko Rieger}
\email{h.rieger@physik.uni-saarland.de}
\affiliation{Theoretische Physik, Saarland University, D-66123 Saarbr{\"u}cken, Germany} 
\author{Ferenc Igl{\'o}i}
\email{igloi.ferenc@wigner.mta.hu}
\affiliation{Wigner Research Centre for Physics, Institute for Solid State Physics and Optics, H-1525 Budapest, P.O. Box 49, Hungary}
\affiliation{Institute of Theoretical Physics, Szeged University, H-6720 Szeged, Hungary}
\date{\today}

\begin{abstract}
We consider the quantum Ising chain with uniformly distributed random antiferromagnetic couplings $(1 \le J_i \le 2)$ and uniformly distributed random transverse fields ($\Gamma_0 \le \Gamma_i \le 2\Gamma_0$) in the presence of a homogeneous longitudinal field, $h$. Using different numerical techniques (DMRG, combinatorial optimisation and strong disorder RG methods) we explore the phase diagram, which consists of an ordered and a disordered phase. At one end of the transition line ($h=0,\Gamma_0=1$) there is an infinite disorder quantum fixed point, while at the other end ($h=2,\Gamma_0=0$) there is a classical random first-order transition point. Close to this fixed point, for $h>2$ and $\Gamma_0>0$ there is a reentrant ordered phase, which is the result of quantum fluctuations by means of an order through disorder phenomenon.

\end{abstract}


\maketitle

\section{Introduction}
\label{sec:intr}
Quantum phase transitions are among the fundamental problems of modern physics, the properties of which are studied in different disciplines: solid state physics, quantum field theory, quantum information, and statistical mechanics\cite{sachdev}. Quantum phase transitions take place at $T = 0$ temperature and these are indicated by singularities in the ground-state expectational values of some observables by varying a control parameter, such as the strength of a transverse field. One basic question in this field of research is how quenched disorder influences the properties of quantum phases and the singularities associated with the quantum phase transitions. This latter problem is theoretically very challenging, since the corresponding quantum state is the result of an interplay between quantum and disorder fluctuations, strong correlations, and frustration.

Many results in this field are known on the random transverse-field Ising chain with short-range interactions. It was Fisher\cite{fisher}, who used a strong disorder renormalization group (SDRG) method\cite{im} and obtained several asymptotically exact results. The phase-transition is shown to be controlled by a so called infinite disorder fixed point\cite{danielreview} (IDFP), at which the distribution of the parameters (couplings and random transverse fields) increase without limit during renormalization. Outside the quantum critical point dynamical observables (susceptibility, autocorrelation functions, etc.) are still singular, due to the presence of strong Griffiths singularities, which are the result of rare regions in the disordered samples\cite{ijl01,vojta}. 

IDFP properties are found also for random Heisenberg chains\cite{fisher_XX,hyman_yang,cecile} as well as in the random singlet phases of $SU(2)_k$ anyonic chains\cite{refael09}. In higher dimensional systems with discrete symmetry the presence of IDFP-s has been demonstrated by numerical application of the SDRG method\cite{2d,2dRG,ddRG}, as well as through quantum MC simulations\cite{QMC1,QMC2}. On the contrary the phase-transition of the random transverse-field Ising model with long-range interactions, the strength of which decays as a power with the distance is shown to have a conventional random quantum fixed point\cite{juhasz14,kovacs16}. The RG flow in this case is similar to that of disordered bosons\cite{altman04,altman10}.

A quantum system is often described by several parameters and in their space the phase-transition takes place at a line or at a (higher dimensional) surface. The phase transition of the clean system in this case is generally governed by a few fixed points, or in exceptional cases by a line of fixed points. In some cases (such as in the random quantum Ashkin-Teller chain\cite{carlon_lajko_igloi,barghathi15}) the clean fixed points are found to turn to the same type of IDFP-s, mainly due to symmetry reasons. However, there are no detailed studies in the general cases, when the different clean fixed points transform differently due to disorder.

In this paper we are going to study such a more complex system, the antiferromagnetic Ising chain in a mixed transverse and longitudinal field. The clean model has two fixed points. The quantum fixed point governs the critical behaviour at any non-zero transverse field, while at zero transverse field there is a classical fixed point, which describes a first-order transition. In the present study we are going to keep the longitudinal fields non-random, but at the same time the couplings and the transverse fields disordered and investigate the phase-transition properties of this random system. We explore the phase-diagram numerically by the DMRG method for finite values of the transverse field, while in the zero transverse-field limit, when the system is classical we use combinatorial optimisation methods to find the true ground state configuration. We also use approximate SDRG calculations, as well as perturbation calculations to see the stability of the random fixed-points. 

The rest of the paper organised in the following way. In Sec.\ref{sec:model} the model is introduced and its properties are described for non-random parameters. The disordered model is studied in Sec.\ref{sec:disordered}. The two end-points of the transition line (zero longitudinal field and zero transverse field) are described asymptotically exactly, while the complete phase-diagram is explored numerically through DMRG and SDRG methods. Our results are discussed in Sec.\ref{sec:disc} and some detailed calculations are presented in the Appendices.

\section{The model}
\label{sec:model}

\begin{figure}[h!]
\begin{center}
\includegraphics[width=8.6cm]{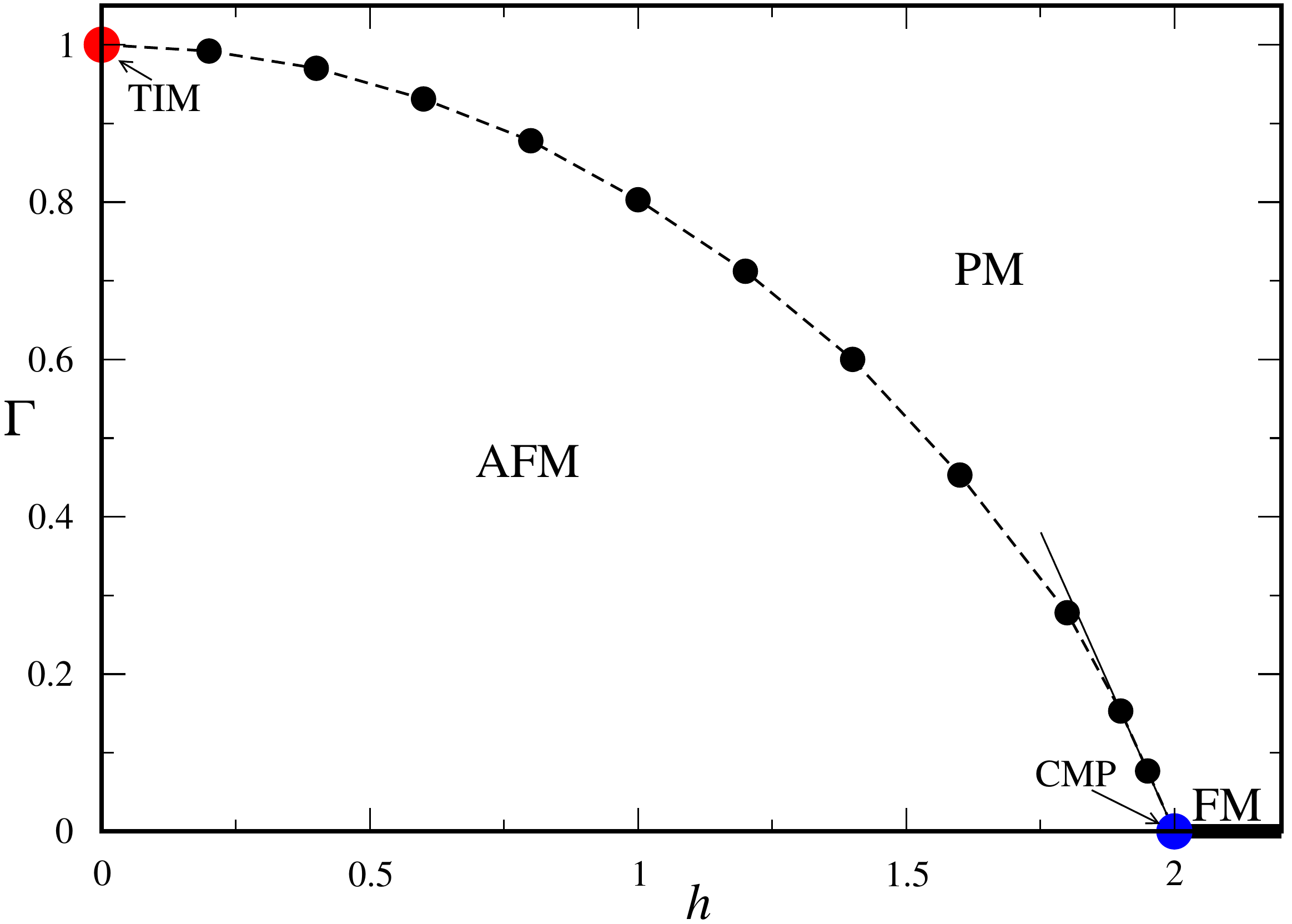}
\end{center}
\caption{\label{fig:clean_phase}(Color online) The zero-temperature phase diagram of the clean antiferromagnetic Ising chain with $J=1$
in transverse ($\Gamma$) and longitudinal ($h$) magnetic fields calculated by DMRG method. 
The transition between a quantum antiferromagnetic (AFM) phase and a quantum paramagnetic (PM) phase
is controlled by the fixed point of the transverse Ising model (TIM) at $(\Gamma=1,\,h=0)$. The classical multicritical point (CMP) at $(\Gamma=0,\,h=2)$ 
corresponds to a first-order transition between an AFM phase and a ferromagnetic (FM) phase
in the absence of quantum fluctuations.  
Near the classical multicritical point the phase-boundary is linear\cite{hard_rods}: $\Gamma_c\approx 1.526492(2-h)$.
}
\end{figure}

We consider the antiferromagnetic Ising chain with mixed transverse and longitudinal fields. The longitudinal fields are constant, $h \ge 0$, while the nearest neighbour couplings, $J_i>0$, and the transverse fields, $\Gamma_i>0$, may be random, so that the Hamiltonian is defined as:
\begin{align}
\begin{split}
\hat{H}&=\sum_{i=1}^L J_i \sigma_{i}^{z} \sigma_{i+1}^{z}\\
&-\sum_{i=1}^L \Gamma_i \sigma^x_{i} -h\sum_{i=1}^L \sigma^z_{i}\;.
\label{Hamilton}
\end{split}
\end{align}
in terms of the $\sigma_{i}^{x,z}$ Pauli matrices at site $i$. For finite chains we have $L=4 \ell$ lattice sites and periodic boundary conditions (PBC-s).

To the best of our knowledge (the clean version of) this model was studied in the second part of the eighties. A more general model, with a $m$-spin product interaction term, i.e. $\prod_{j=0}^{m-1}\sigma_{i+j}^{z}$ instead of $\sigma_{i}^{z} \sigma_{i+1}^{z}$ has been introduced and studied for $m=3$ by Penson \textit{et al.}\cite{penson} and for $h>0$ a phase transition of the 3-state Potts universality class has been found. Soon after this model is considered in the vicinity of the classical limit $\Gamma \to 0$ and $h \to m$ (which is called as quantum hard rods) and has been studied by finite-size exact diagonalization\cite{hard_rods}. For dimers with $m=2$ the transition is shown to be in the quantum Ising universality class, while for $m=3$ the transition is in the 3-state Potts universality class, which has been shown more precisely by MC simulations on the classical version of the model\cite{rods_MC1,rods_MC2}. Finally a detailed study of the clean model in Eq.(\ref{Hamilton}) has been performed in\cite{ovchinnikov} after a preliminary investigation in\cite{sen}.

\subsection{Clean model}

First we consider the clean model with $J_i=1$ and magnetic fields $\Gamma_j=\Gamma$ and $h$ and the phase diagram is shown in Fig.\ref{fig:clean_phase} which has been calculated by DMRG method. This phase-diagram agrees with the previous calculations obtained by DMRG\cite{ovchinnikov}, by experimental realization in an optical lattice\cite{simon}, by quantum MC\cite{Yu_Cheng} and by the fidelity susceptibility method\cite{bonfim}. The phase diagram in Fig.\ref{fig:clean_phase} for finite $\Gamma>0$ contains an ordered antiferromagnetic phase (AFM) and a paramagnetic phase (PM). The transition between them is controlled by a quantum Ising fixed-point at $(\Gamma=1,\,h=0)$, the properties of which are known exactly\cite{pfeuty}. 

The transition line ends at $(\Gamma=0,\,h=2)$, where there is a classical multicritical point (CMP). In the limiting case $\Gamma \to 0$ and $h \to 2$ the system reduces to the quantum hard dimer model, having the transition at\cite{hard_rods} $\Gamma=1.526492(2-h)$. This is to be compared with direct calculations $\Gamma \approx 1.5 (2-h)$ in\cite{ovchinnikov} and $\Gamma \approx 1.4 (2-h)$ in\cite{Yu_Cheng}. The quantum Ising nature of the transition in the hard dimer limit has been performed with large numerical precision, as well as the conformal properties have been determined. At $\Gamma=0$ the system is classical and there are no quantum fluctuations. Here the transition takes place at the classical multicritical point (CMP) which is of first order between the AFM phase and a ferromagnetic (FM) phase. At the CMP the ground state is infinitely degenerate, the entropy per site is finite\cite{domb}.

\section{Disordered model}
\label{sec:disordered}

Here we consider random variables in Eq.(\ref{Hamilton}), so that the antiferromagnetic couplings and the transverse fields are independent random numbers, which are taken from some distributions. In the numerical calculations we used box-like distributions:
\begin{align}
  \begin{split}
    \pi_1(J) &=
    \begin{cases}
      1 & \hspace*{0.65cm}\text{for } J_0<J\le J_0+1\,,\\
      0 & \hspace*{0.65cm}\text{otherwise.}
    \end{cases} \\
    \pi_2(\Gamma) &= 
    \begin{cases}
      1/\Gamma_0 & \text{for } J_0\Gamma_0<\Gamma \le (J_0+1)\Gamma_0\,,\\
      0 & \text{otherwise.} 
    \end{cases}
  \end{split} 
  \label{eq:J_distrib} 
\end{align}
In the following we argue, that the phase-diagram of the random system is different, if the smallest coupling is $J_{min}=J_0>0$ (when there is an extended ordered region in $h$) or $J_{min}=J_0=0$ (when the ordered region is restricted to $h=0$). Indeed, in the classical limit with $\Gamma_0=0$ the ground state is strictly AFM ordered, if $h<2 J_{min}$ and for $h>2 J_{min}$ domain-wall excitations destroy the AFM order (see in more details in Sec.\ref{sec:classical}). In the following we shall investigate the region with $J_0>0$ and in the numerical work we keep $J_0=1$. We note, that the case $J_0=0$ has been studied earlier by Lin \textit{et al}\cite{Yu_Cheng}, so that we are going to study an unexplored system.

Now let us have a look at the phase-diagram of the clean system in Fig.\ref{fig:clean_phase} having two fixed points, which are located at the two ends of the phase-transition line. In the following we study their stability with respect to disorder in Eq.(\ref{eq:J_distrib}). The TIM fixed-point at $h=0$ is transformed to the random transverse Ising model (RTIM), for which several asymptotically exact results are known through SDRG calculations, which are shortly collected in Sec.\ref{sec:h=0} . The other fixed point, the classical multicritical point at $\Gamma=0$ is transformed to a random classical multicritical point (RCMP), the properties of which are studied in Sec.\ref{sec:classical}.

\subsection{RTIM at $h=0$}
\label{sec:h=0}

For zero longitudinal field, $h=0$, the model is equivalent to the random transverse Ising model, for which many asymptotically exact results are known due to SDRG calculations~\cite{fisher}, what we shortly recapitulate here. The critical point is located at:
\be
    \overline{\ln J}=\overline{\ln \Gamma}\,.
    \label{eq:duality}
\ee
thus with the distribution in Eq.(\ref{eq:J_distrib}) we have $\Gamma_0^c(h=0)=1$.

The energy-scale in the system is the excitation energy (smallest gap) $\epsilon$ and its relation with the length-scale, $L$, at the critical point is given by:
\be
\ln \epsilon \sim L^{\psi}
\label{psi}
\ee
with a critical exponent $\psi=1/2$.

The spin-spin correlations are defined as: 
\be
C(r)=(-1)^r \langle \sigma_i^z\sigma_{i+r}^z \rangle 
\label{C(r)}
\ee
and their average decay is given as a power of $r$ at the critical point,
\be
      \overline{C}(r)\sim \frac{1}{r^{2-\phi}}\,,
\ee
where
$ \phi=(1+\sqrt{5})/2$
is the golden mean.

The deviation from criticality is parameterised by~\cite{fisher}
\be
 \delta=\frac{\overline{\ln \Gamma}-\overline{\ln J}}{\text{var}(\ln h)+\text{var}(\ln J)}\,,    
 \label{eq:delta}
\ee
where $\text{var}(x)$ stands for the variance of the random variable $x$.
In the disordered phase $\delta>0$, the average correlations 
decay exponentially with the true correlation length $\xi \sim 1/\delta^2$, implying the correlation length exponent $\nu=2$ for the random chain.

Outside the critical point the relation between the energy- and the legth-scale is given by:
\be
\epsilon \sim L^{-z}
\label{L_z}
\ee
where the dynamical exponent, $z$ is given by the positive root of the equation:
\be
      \overline{\left(\frac{J}{\Gamma} \right)^{1/z}}=1\,,
      \label{eq:z_eq}
\ee
which is an exact expression in the entire Griffiths region.~\cite{Z,ijl01,Igloi}
In the vicinity of  the critical point, the dynamical exponent to the leading order is given by~\cite{fisher,Igloi}
\be
       z\approx \frac{1}{2|\delta|},\qquad |\delta| \ll 1\,.
\ee 

\subsection{The classical limit: $\Gamma_0=0$}
\label{sec:classical}

\begin{figure}[h!]
\begin{center}
\vskip 0cm
\includegraphics[width=8.cm,angle=0]{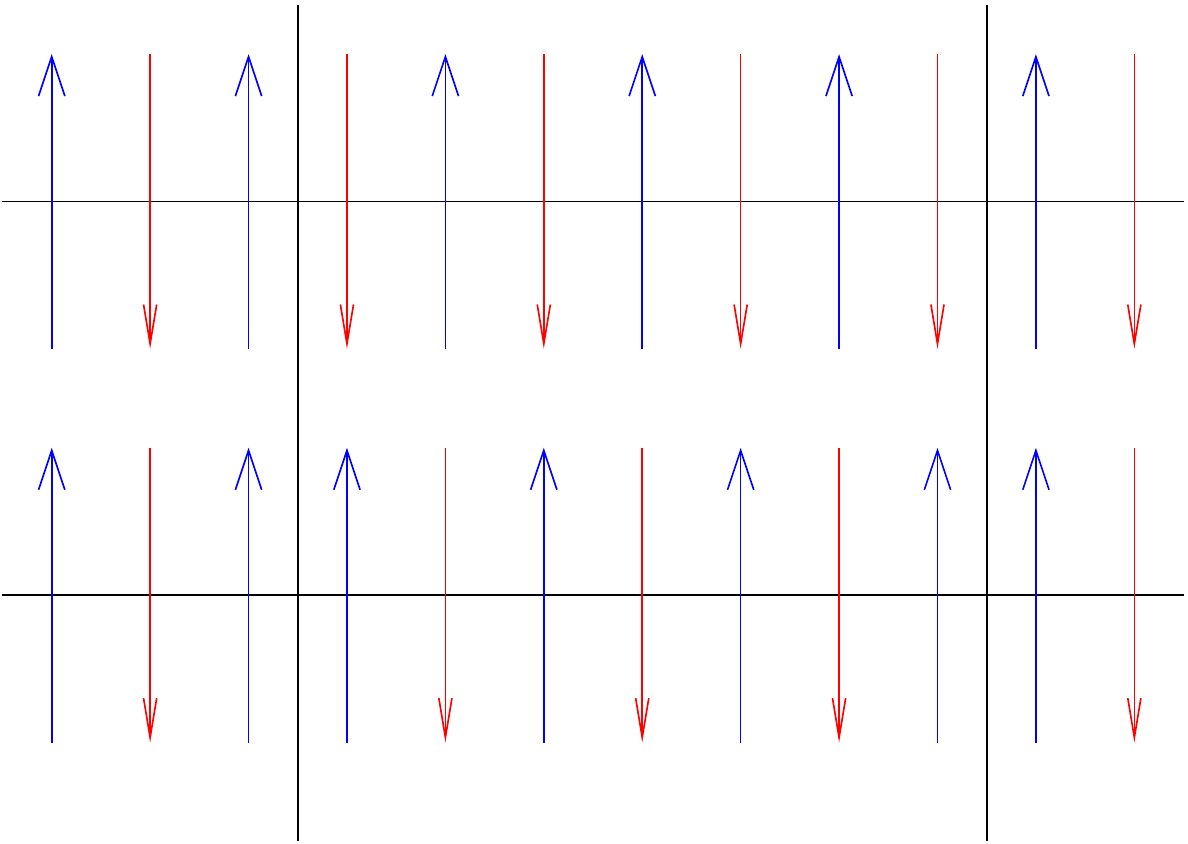}
\end{center}
\vskip 0cm
\caption{\label{fig:domain}(Color online)Illustration of the AFM ground state for $h<J_{i_1}+J_{i_2}$ (first row) and the ground state with two AFM domains for $h>(J_{i_1}+J_{i_2})$ (second row). Here $J_{i_1}$ ($J_{i_2}$) is the smallest random coupling at odd (even) positions. The boundary of the domains are denoted by vertical lines, at $i_1$ and $i_2$, which are in different parity positions.}
\end{figure}

To explore the ground state of the system in the classical limit let us first sort the random couplings in increasing order, separately at odd $1 < J_{i_1} < J_{i_3} < J_{i_5} \dots < J_{i_{L-1}} < 2$ and and even positions: $1 < J_{i_2} < J_{i_4} < J_{i_6} \dots < J_{i_{L}} < 2$ and let us increase gradually the strength of the longitudinal field from $h=0$. For $h< J_{i_1}+J_{i_2}$ the ground state is fully antiferromagnetic, since the longitudinal field is too weak to create a turned domain. This is illustrated in the first row of Fig.\ref{fig:domain}. For $h> J_{i_1}+J_{i_2}$, however, it is energetically favourable to create domain walls at $i_1$ and $i_2$ and thus turn the spins in the domain between $i_1$ and $i_2$. At the boundary of the domains the spins are in $\uparrow \uparrow$ positions. This is illustrated in the second row of Fig.\ref{fig:domain}. By increasing the value of $h$ further more and more domains and thus domain walls are created and consequently the FM order monotonously increasing. Passing $h=4$ the system is fully FM ordered.

\begin{figure}[h!]
\begin{center}
\hskip 0cm
\includegraphics[width=10.cm,angle=0]{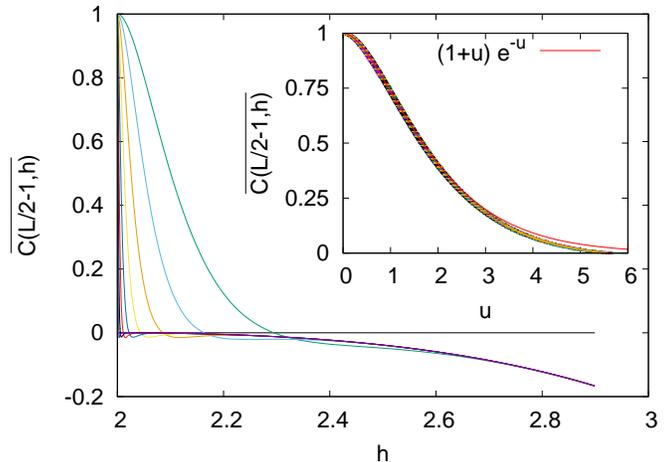}
\end{center}
\vskip -0cm
\caption{\label{fig:G=0}(Color online) The average correlation function as a function of $h$ for different sizes of the system. From right to left $L=2^5,2^6,2^7,\dots, 2^{17}$. The effective transition point, $h^*_L$, is defined as the position of the zero point. Inset: scaling plot in terms of $u=(h-2)L/2$. The results for $L=2^7,2^8,2^9,\dots, 2^{17}$ are indistinguishable. The red line represents the theoretical estimate obtained in the case when the ground state  contains only one or two domains.
}
\end{figure}

Now let us concentrate on the properties of the system close to $h=2$ where the AFM order is lost. The AFM long-range order in the system is characterised by the average correlation function in Eq.(\ref{C(r)}), and we choose the two points of reference to be separated by the maximal distance, $r=L/2-1$. We have calculated $\overline{C(L/2-1,h)}$ numerically by an algorithm, which is described in the Appendix \ref{app:algorithm}. We obtained the ground state of periodic samples of sizes $L=2^n$, up to $n=17$, thus the largest samples have a length $L=131072$. The typical number of random realisations were about thousand, for not extreme large samples.
The field dependence of the average correlation function is shown in Fig.\ref{fig:G=0} for different sizes. It is seen, that there is a transition regime from AFM order with $\overline{C(L/2-1,h)}>0$ to the FM regime with $\overline{C(L/2-1,h)}<0$, having a width $\Delta h_L$ which is more and more sharp as the size of the system increases and at the same time the position of an effective transition point, $h^*_L$, defined as the position of the zero-point (or the inflection point) of the curve approaches $h=2$. From the numerical results we draw the conclusion, that for the box-distribution in Eq.(\ref{eq:J_distrib}) what we used here we have the relations: $\Delta h_L \sim [h^*_L-2] \sim 1/L$. This means that the appropriate scaling variable in the transition regime is $u=(h-2)L/2$.\cite{omega}
Indeed in terms of $u$ the average correlation functions in the transition regime collapse on one master curve as illustrated in the inset of Fig.\ref{fig:G=0}. This master curve, at least for small values of $u$ can be calculated exactly. If we restrict ourselves to such samples which are either fully AFM ordered or consists of just two domains $\overline{C(L/2-1,h)}$ can be calculated through extreme value statistics\cite{galambos,ev}. As described in detail in the Appendix \ref{app:corr_func}, the average correlation function in such an approximation is given by:
\be
\overline{C(L/2-1,h)} \approx (1+u) \exp(-u),\quad u \ll 1\;.
\label{C_appox}
\ee
As shown in the inset of Fig.\ref{fig:G=0} this function describes very well the average correlation function, even for not too small values of $u$.
In the thermodynamic limit the transition takes place at $h=2$, where the average correlation function exhibits a jump: $\lim_{h \to 2^-} lim_{L \to \infty}\langle C(L/2-1,h)\rangle=1$ and $\lim_{h \to 2^+} lim_{L \to \infty}\langle C(L/2-1,h)\rangle=0$, thus we have a random first-order transition. We should note, however, that for $h>h^*=2$ there is a divergent length-scale: $\xi \sim (h-h^*)^{-1}$, which measures the typical size of the AFM domains in the system. In this respect the transition is of mixed order\cite{bar_mukamel}.

\subsection{DMRG results for $h>0$}
\label{sec:dmrg}

The complete phase-diagram of the model has been studied numerically by the DMRG method. In this investigation we used finite samples of length $L=4,8,16,32$ and $64$ and their ground state and the first few excited states are calculated. Here the original version of the infinite-size DMRG scheme was utilised for with PBC\cite{dmrg1,dmrg2}. The accuracy of the ground-state energy calculations was in the range of $10^{-6}$–$10^{-8}$ and this was in full agreement with the truncation error, the largest basis size being $m = 100–200$ for the different systems. 

Averages are performed typically over a few ten thousand independent samples which have a microcanonical distribution. For the couplings the microcanonical ensemble is obtained from a canonical ensemble $1<J_i<2$, $i=1,2,\dots,L-1$, by calculating their average $\overline{J}=\sum_i^{L-1} J_i /(L-1)$ and transforming the couplings as: $\tilde{J_i}=J_i-\overline{J}+3/2$. Evidently the microcanonical set of couplings, $\tilde{J_i}$ have the same average value, $3/2$, for all random samples. We use a similar rule to define the microcanonical set of random transverse fields. In the thermodynamical limit the canonical and the microcanonical ensambles lead to identical averages. For finite systems, however, the microcanonical values usually have smaller sample to sample fluctuations.
 
We have studied the behaviour of the average correlation function, $\overline{C(L/2-1,h,\Gamma_0)}$, what we have considered also in the classical limit in Sec.\ref{sec:classical}. We have checked, that the DMRG results at $\Gamma_0=0$ agree with those calculated previously by combinatorial optimisation methods. We remind that at $\Gamma_0=0$ an effective, size-dependent transition point, $h^*_L(0)$, has been defined through the position of the zero-value of the average correlation function and we use the same criterion in the quantum regime, for $\Gamma_0>0$, too. 
\begin{figure}[h!]
\begin{center}
\hskip 1cm
\includegraphics[width=8.cm,angle=0]{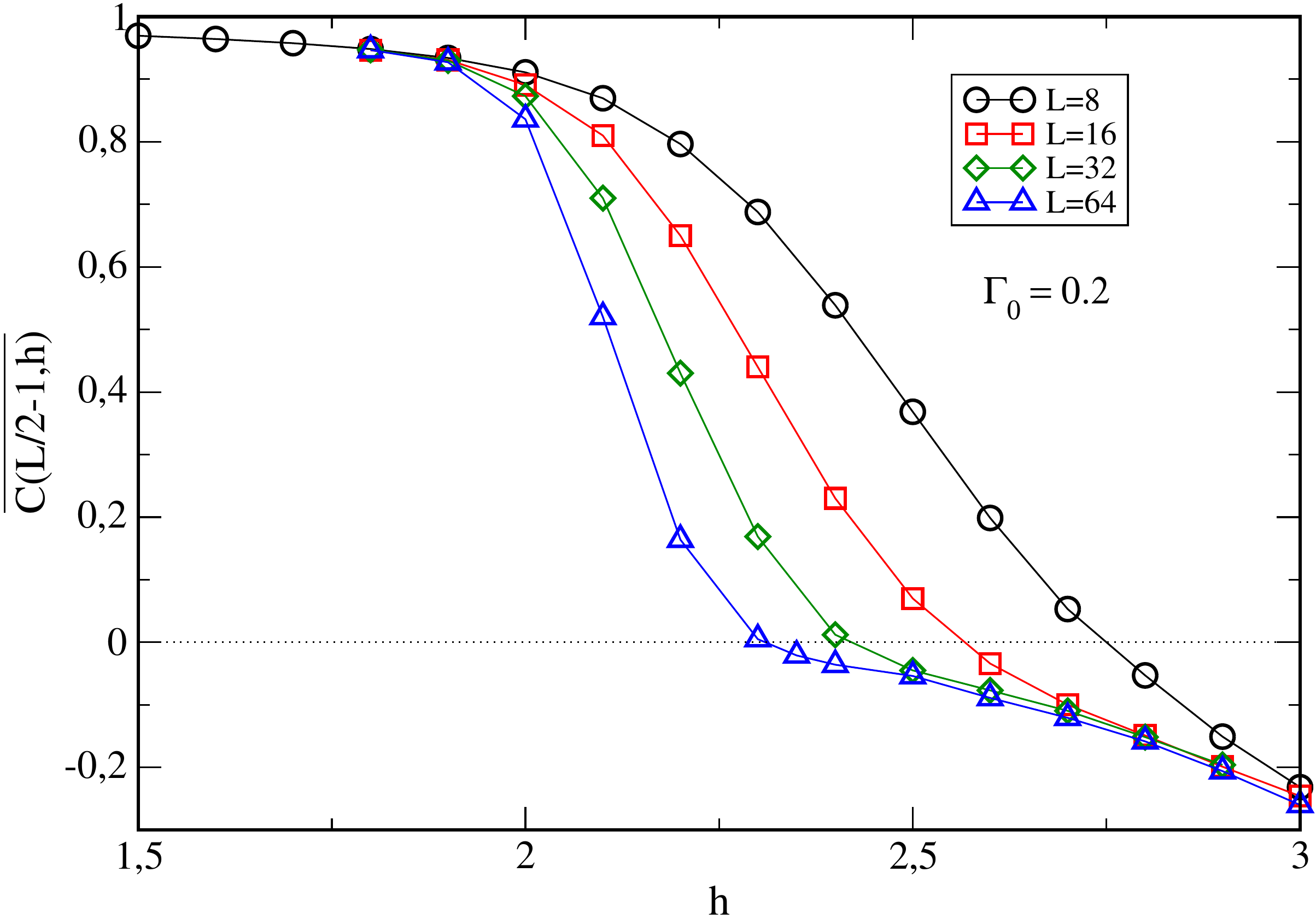}
\end{center}
\caption{\label{fig:gamma_0.2}(Color online) The average correlation function at $\Gamma_0=0.2$ as a function of $h$ for different finite systems. The position of its zero point is used to define effective, size-dependent transition points. These are depicted in Fig.\ref{fig:dmrg_phase} for different values of $\Gamma_0$.}
\end{figure}

This is illustrated in Fig.\ref{fig:gamma_0.2}, where $\overline{C(L/2-1,h,\Gamma_0)}$ is shown for $\Gamma_0=0.2$ as a function of $h$ and for different values of $L$. 

\begin{figure}[h!]
\begin{center}
\hskip 1cm
\includegraphics[width=8.cm,angle=0]{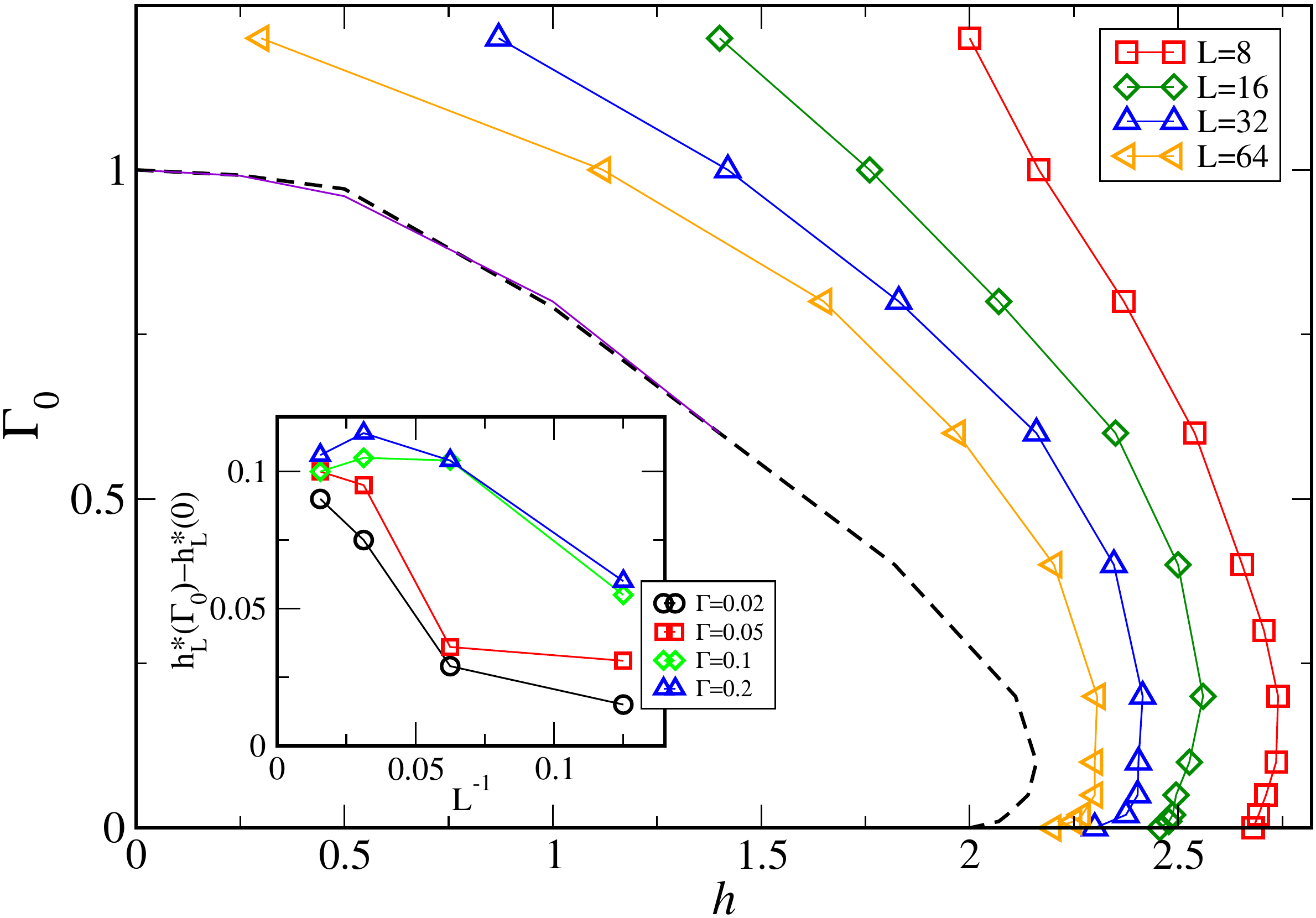}
\end{center}
\caption{\label{fig:dmrg_phase}(Color online) Numerically estimated finite-size transition points calculated by DMRG method for systems with $L=8,16,32$ and $64$. The dashed line is guide to the eye, representing the expected true phase-boundary at the thermodynamical limit. Inset: difference between the effective transition points: $\Delta h^*_L(\Gamma_0)= h^*_L(\Gamma_0)-h^*_L(0)$ as a function of $1/L$ for different values of $\Gamma_0$.
}
\end{figure}
The effective, size-dependent transition points calculated in this way are presented in Fig.\ref{fig:dmrg_phase}. One can notice in this figure, that close to the classical limit the effective transition points start to increase with $\Gamma_0>0$, have their maximum value around $\Gamma_0 \approx 0.2$ and then monotonously decrease as $\Gamma_0$ increases further. Before performing a more detailed analysis we note that at $\Gamma_0$ we know the true transition point, $h^*(0)=2.0$, which is formally the extrapolated value of the series $h^*_L(0)$. We have checked, that in the regime $L \le 64$ there are considerable corrections to scaling contributions and similar behaviour is expected to happen by extrapolating the data for $\Gamma_0>0$, too. Therefore in the vicinity of the classical limit we analyse the difference:
$\Delta h^*_L(\Gamma_0)= h^*_L(\Gamma_0)-h^*_L(0)$, which is plotted in the inset of Fig.\ref{fig:dmrg_phase}. For smaller sizes, $L=8$ and $16$ there are large corrections, in particular for small values of $\Gamma_0$. At the larger sizes ($L=32$ and $64$), however the differences, $\Delta h^*_L(\Gamma_0)$ are stable, and these are positive for $0<\Gamma_0 \le 0.2$. We expect, that this trend remains in the thermodynamic limit, too.

From this assumption follows the conclusion, that in the phase-diagram of the random antiferromagnetic Ising model with mixed transverse and longitudinal fields there is a reentrance behaviour. Starting at $\Gamma_0=0$ and selecting a longitudinal field $h$, which is somewhat larger than the transition point $h^*(0)=2$, the system is disordered. By switching on quantum fluctuations with increasing $\Gamma_0$ there is an order through disorder phenomena, so that the system stays ordered for $\Gamma_0^1<\Gamma_0 < \Gamma_0^2$ and remains disordered for $\Gamma_0>\Gamma_0^2$. The presence of reentrance in our system can be shown exactly by calculating the quantum corrections to the classical limit in the vicinity of the RCMP. This is shown in the Appendix \ref{sec:Q_corrections}.

\begin{figure}[h!]
\begin{center}
\hskip 1cm
\includegraphics[width=9.cm,angle=0]{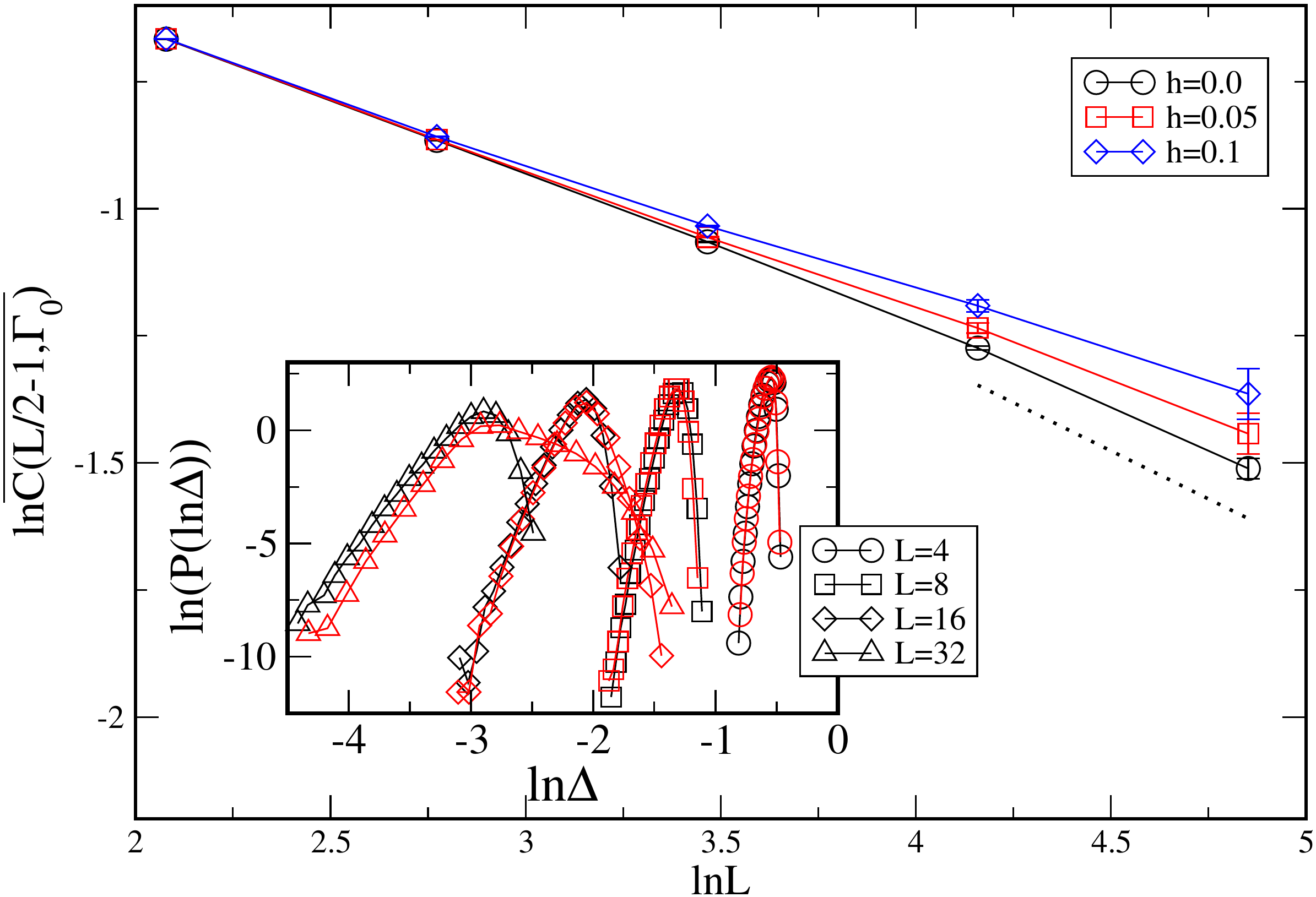}
\end{center}
\caption{\label{fig:gap}(Color online) Average critical correlation function, $\overline{C(L/2-1,\Gamma_c)}$ as a function of $L$ in log-log plot for different values of $h=0.0,0.05$ and $0.1$. The dotted line has the slope: $2-d_f=0.38$, corresponding to the exact result of the decay exponent at $h=0.0$, see in Eq.(\ref{C(r)}). Inset: distribution of the excitation gaps at $h=0.2$ and $\Gamma_0=0.992$ at different finite systems (red symbols) and compared with those at the RTIM point with $h=0.0$ (black symbols)}.
\end{figure}

At the other end of the phase-transition line, close to the RTIM fixed point at $h=0$ the analysis of the average correlation function leads to less accurate results. In this regime we estimated the location of the transition line in the followig way. We calculated the distribution of the excitation gaps at finite chains and compared those with the same type of distributions at the RTIM. Having the gap-distributions at $h>0$ for different values of $\Gamma_0$ the transition point is identified, with that value of $\Gamma_0$, where the (small-gap part of the) gap-distributions were the closer to that at the RTIM at the same size. We illustrate this procedure in the inset of Fig.\ref{fig:gap}. Since the distributions are found very similar to that in the RTIM we conclude that for $h>0$ there is infinite disorder scaling, so that the scaling behaviour of the small gaps is well described by the relation in Eq.(\ref{psi}) and the scaling exponent is (very close to) $\psi=1/2$. 

In the next step we calculate the critical average correlation functions, where the transition points have been estimated in the previous paragraph from the scaling of the gaps. From a log-log plot of $\overline{C(L/2-1,\Gamma_c)}$ vs. $L$ in Fig.\ref{fig:gap} one can notice an asymptotic linear trend, which corresponds to a power-low dependence. At $h=0$ the slope of the lines is compatible with the exact result of the decay exponent in Eq.(\ref{C(r)}). For somewhat larger longitudinal fields, $h=0.05$ and $0.1$, the slopes of the line segments are somewhat larger than at $h=0.0$, but due to errors of the calculation one can not exclude the possibility, that the asymptotic exponents agree with that at $h=0.0$. 

We shall come back to study the scaling behavior of the system for small values of $h$ by the SDRG method in the following Sec.\ref{sec:SDRG} and postpone the analysis of the phase-diagram to the discussion in Sec.\ref{sec:disc} .

\subsection{SDRG calculations for $h \ll 1$}
\label{sec:SDRG}

Here we are going to extend the SDRG method\cite{fisher}, which has been developed for $h=0$ by including small longitudinal fields which are treated as a perturbation. For the sake of simplicity we use here the transformation $\sigma^z_j=(-1)^j \sigma^z_j$, when the model is described by random ferromagnetic couplings (and transverse fields) in the presence of a staggered longitudinal field. This last part of the Hamiltonian is generally written as:
\begin{equation}
\hat{H}_{\rm longi}= -\sum_{j=1}^L h_j \sigma^z_{j}\;,
\end{equation}
and in the starting situation $h_j=(-1)^j h$.

In the SDRG procedure we consider the separated local degrees of freedom say at position $i$. These are couplings or sites, having the value of the largest gaps: $2J_i$ and $2\sqrt{\Gamma_i^2+h_i^2}$, respectively. These gaps are sorted in descending order and the largest one, denoted by $\Omega$, which sets the energy-scale in the problem, is eliminated and between remaining degrees of freedom new terms in the Hamiltonian are generated through perturbation calculation. This procedure is successively iterated, during which $\Omega$ is monotonously decreasing. At the fixed point, with $\Omega^*=0$
one makes an analysis of the distribution of the different parameters and calculates the scaling properties. In the following we describe the elementary decimation steps.

\subsubsection{Strong-coupling decimation}

In this case the largest local term in the Hamiltonian is a coupling, say $\Omega=J_i$, connecting sites $i$ and $i+1$ and  the two-site Hamiltonian is given by:
\begin{equation}
\hat{H}_{cp}=-J_i \sigma^z_{i}\sigma^z_{i+1}-\Gamma_i\sigma^x_{i} -\Gamma_{i+1}\sigma^x_{i+1} -(h_i \sigma^z_{i}+h_{i+1} \sigma^z_{i+1})\;.
\label{strong-coupling}
\end{equation}
The spectrum of $\hat{H}_{cp}$ contains four levels, the lower two being separated from higher two by a gap $2J_i$. We omit the higher two levels, which is equivalent to merge the two strongly coupled sites into a spin cluster in the presence of a (renormalized) transverse field $\tilde{\Gamma}$ and a longitudinal field $\tilde{h}$. The magnetic moment of the cluster is given by: $\tilde{\mu}=\mu_i+\mu_{i+1}$, where the magnetic moments in the starting situation are $\mu_i=\mu_{i+1}=1$. The remaining two lowest energy levels are given by second-order degenerate perturbation method as the eigenvalues of the matrix:
\begin{equation}
\begin{bmatrix}
 -J_i-h_i-h_{i+1}       & -\Gamma_i\Gamma_{i+1}/J_i \\
 -\Gamma_i\Gamma_{i+1}/J_i &  -J_i+h_i+h_{i+1} 
\end{bmatrix}
\end{equation}
having a gap 
\be
\Delta E_{cp}=2 \sqrt{\left(\Gamma_i\Gamma_{i+1}/J_i\right)^2+\left(h_i+h_{i+1}\right)^2}\;.
\ee
Comparing it with the gap of the spin cluster we obtain for the renormalised values of the parameters:
\be
\tilde{\Gamma}=\frac{\Gamma_i\Gamma_{i+1}}{J_i},\quad \tilde{h}=h_i+h_{i+1}\;.
\ee
Note, that in the starting situation with $h_i=-h_{i+1}$ after decimating a strong coupling the longitudinal field is eliminated at the effective composite spin. 

\subsubsection{Strong-transverse-field decimation}

In this case the largest local term is a transverse field, say $\Gamma_i$ and the corresponding energy-gap of the one-site Hamiltonian is $2\sqrt{\Gamma_i^2+h_i^2}$. Due to the large $\Gamma_i$ this site does not contribute to the longitudinal magnetisation and therefore it is eliminated. The longitudinal magnetic field, however should be transformed at the remaining neighbouring sites. To calculate the new renormalized coupling between the remaining sites $i-1$ and $i+1$ we calculate energy levels with fixed spins at these sites. Denoting by $s_{i\pm1}=+$ ($-$) a $\uparrow$ ($\downarrow$) boundary state, the eigenvalue problem with different boundary conditions has the lowest energy as:
\begin{equation}
E_{s_i,s_{i+1}}=-\sqrt{\Gamma_i^2+(s_{i-1}J_{i-1}+s_{i+1}J_i+h_i)^2}\;.
\end{equation}
The renormalised coupling between the remaining sites is given by:
\begin{equation}
\tilde{J}=(E_{\uparrow \uparrow}+E_{\downarrow \downarrow}-E_{\uparrow \downarrow}-E_{\downarrow \uparrow})/4 \approx \frac{J_{i-1}J_i}{\sqrt{\Gamma_i^2+h_i^2}}\;,
\end{equation}
where the last relation is calculated perturbatively.

%
%
%
%

Concerning renormalization of the longitudinal magnetic fields we require that the sum of these fields is locally conserved, in agreement with the original Hamiltonian. This is obtained by adding $h_i/2$ to the longitudinal fields at the neighbouring sites: $\tilde{h}_{i\pm1}={h}_{i\pm1}+h_i/2$ and in this way we avoid random-field effects.

\subsubsection{Numerical iteration of the SDRG equations}

We have iterated the decimation equations presented in the previous sections for finite periodic chains of length $L=256,512$ and $1024$ up to the last pair of spins and the energy-gap of this dimer is identified as the gap of the given sample. We have considered $10000$ independent samples for each cases. We have also calculated the total magnetic moment of the samples, and calculated their average value: $\mu_L$, which scales differently in the different phases. In the ordered phase it is extensive, $\mu_L \sim L$, while in the disordered phase it approaches a finite limiting value. At the transition point there is a power-law dependence: $\mu_L \sim L^{d_f}$, with a fractal dimension $0<d_f<1$. This is related to the decay exponent of the correlation function, since $\overline{C(L/2-1,\Gamma_c)} \sim L^{-2(1-d_f)}$, for large $L$.

\begin{figure}[h!]
\begin{center}
\hskip 1cm
\includegraphics[width=9.cm,angle=0]{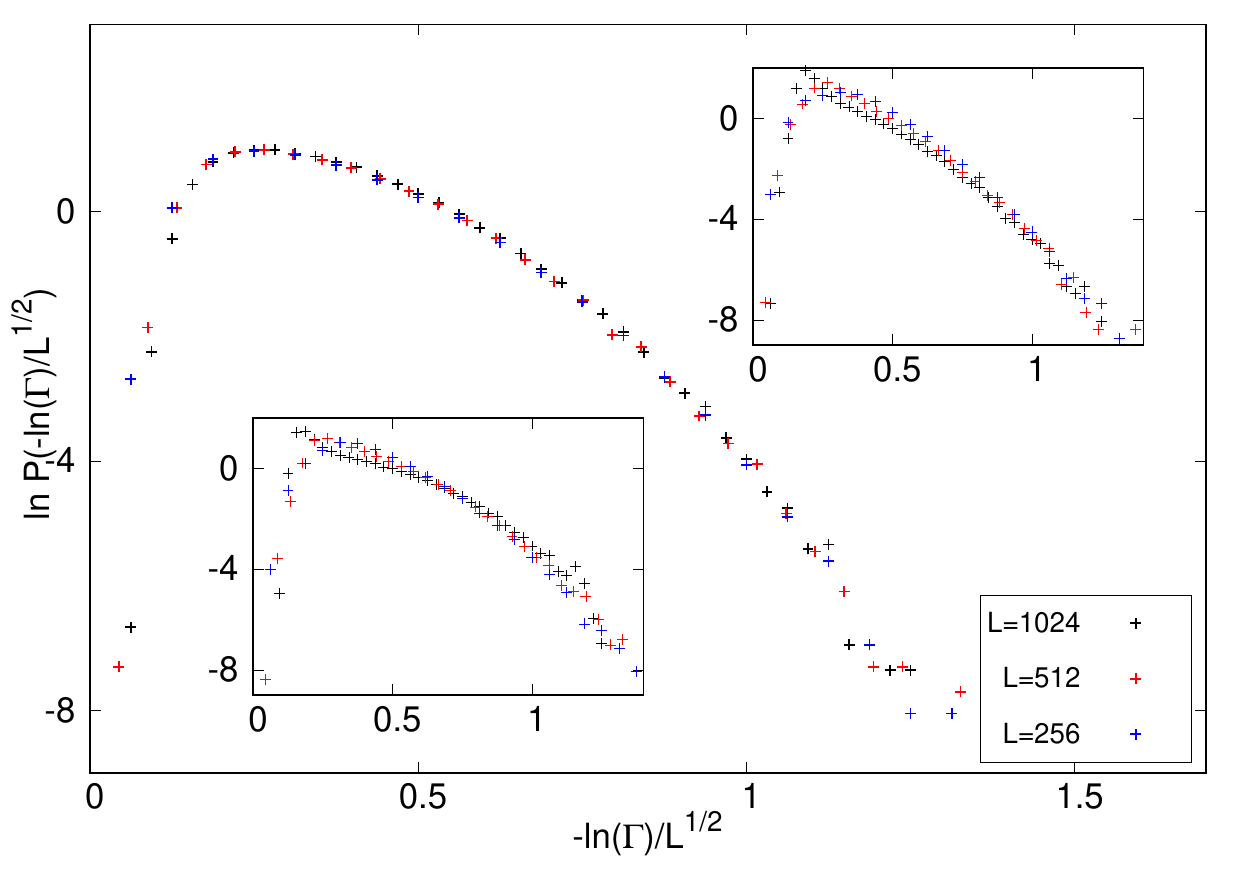}
\end{center}
\caption{\label{fig:gap_distr}(Color online) Scaling plot of the distribution of the energy gaps calculated by the SDRG algorithm at the critical point of the model for finite periodic chains of lengths $L=256,512$ and $1024$. $h=0.$ main panel, $h=0.05$ right inset, $h=0.1$ left inset.}
\end{figure}

First, we have checked that at the RTIM fixed-point with $h=0.0$ the critical point is at $\Gamma_0^*=1.$ and the magnetic fractal dimension is $d_f=0.81$, which is in good agreement with the analytical result: $d_f=\phi/2$, see in Eq.(\ref{C(r)}). The distribution of the gaps, as shown in the main panel of Fig.\ref{fig:gap_distr} is also in agreement with the scaling relation in Eq.(\ref{psi}).

Next switch on a small longitudinal field, $h=0.05$, when the transition point is moved to $\Gamma_0^* \approx 0.9975$. The estimate for the magnetic fractal dimension, $d_f \approx 0.83$ is somewhat larger, than at $h=0$. The calculated gaps show infinite disorder scaling in the right inset of Fig.\ref{fig:gap_distr} with an exponent $\psi \approx 1/2$. The scaling collapse of the data points in this case is less proper for larger values of the gap, but the scaling for the small-energy tail of the distribution is convincing.

Repeating the calculation with a somewhat larger longitudinal field, $h=0.1$, the transition point is shifted further to:
$\Gamma_0^* \approx 0.99$. This means that for small value of $h$ the transition point has a quadratic dependence: $\Gamma_0^* -1. \sim h^2$, like in the clean system. The estimate for the magnetic scaling dimension has grown further,
$d_f \approx 0.89$, while the gaps in the left inset of Fig.\ref{fig:gap_distr} show infinite disorder scaling, in a similar way, as for $h=0.05$.

If we increase the value of $h$ further, the SDRG iteration shows numerical problems. There is a fraction of samples in which the small $h$ condition does not work any more: at some steps the largest term in the Hamiltonian can be a longitudinal field and at that point the SDRG algorithm breaks down.

\section{Discussion}
\label{sec:disc}

We have studied the phase diagram of the antiferromagnetic Ising chain with random couplings and random transverse fields in the presence of homogeneous longitudinal fields. The distribution of the couplings in our study has a finite limiting lower value, $J_i>J_0>0$. In this case the system exhibits an AFM ordered phase and a disordered phase, which are separated by a transition line. The expected form of the phase-transition line is shown in Fig.\ref{fig:dmrg_phase}: it starts at $h=0, \Gamma_0=1$ in the RTIM fixed point and ends at $h=2, \Gamma_0=0$ at the RCMP point. The phase transitions at the two endpoints are completely different: in the RTIM fixed point there is infinite disorder scaling, while in the RCMP the transition is of random first order. Switching on a homogeneous longitudinal field at the RTIM the infinite-disorder scaling of the transition remains valid, which has been shown by SDRG calculations and by DMRG results. Along the transition line for small $h$ the energy-scaling exponent, $\psi=1/2$, seems to be constant, while the fractal dimension of the magnetisation, $d_f$, shows some $h$-dependence.

In the vicinity of the RCMP by including quantum fluctuations, $\Gamma_0>0$, the system shows reentrant behaviour: for $h>2$ the system
with increasing value of $\Gamma_0$ first moves from the disordered phase to the AFM phase and then back to the disordered one.
Reentrance, to our best knowledge has not been observed in random quantum systems so far. In classical systems it is usually the result of competing interactions and/or frustration\cite{SG}. In our system at the clean classical multicritical point, MCP the ground state is infinitely degenerate. This degeneracy is lifted through disorder\cite{villain} at several steps, but the new, non-ordered states are typically less favoured by quantum fluctuations, than the ordered state, which leads to the reentrant behaviour.

Between the infinite disorder scaling regime and the random first-order transition region there must by a repulsive, multicritical fixed point, which separates the two parts of the transition line. This fixed point is expected to be the result of the competition between random couplings, random transverse fields and the homogeneous longitudinal field. With our present investigations we could not explore the properties of this hypothetical fixed point, it will be the aim of further studies. 

Outside the transition line the system exhibits singular dynamical behaviour due to Griffiths singularities. Near the RTIM these are similar to that mentioned in Sec.\ref{sec:h=0}, see in Eqs.(\ref{L_z}) and (\ref{eq:z_eq}). These have been studied in more details in\cite{Yu_Cheng}.

The model can be extended and generalised in different directions. In higher dimensions one should consider bipartite lattices, which can accommodate AFM order. Here, at $h=0$ there is a higher dimensional RTIM fixed-pont, which is known to be infinite disorder type\cite{2d,2dRG,ddRG}. Here one should study first the behaviour in the classical limit and then the complete phase-transition line. Finally, one can also extend our model with random longitudinal fields. In this case, however the disorder fluctuation are so strong, that no ordered phase exits, even with vanishing quantum fluctuations.

\appendix
\section{Numerical algorithm to calculate the ground state in the classical limit}
\label{app:algorithm}

It is possible to map the problem of finding the ground-state onto a max-flow
problem \cite{noa}. However, in the one dimensional problem
it is more efficient to use the following simpler algorithm: i) consider the
set $S$ of all pairs of bonds separated by an even number of bonds,
ii) assign to each pair of $S$ the sum of the two couplings, $J_i+J_j$ values, and sort $S$
according to this sum, iii) take each pair $(i,j)$ of $S$ in increasing order
and choose the state with the lowest energy between the three following cases. Starting with an AFM state and 
1) flip all spins between $i$ and $j$; 2) flip all spins between $j$ and $i$ or 3) flip no spin. Note that we assume $L$ even, and periodic boundary conditions. This algorithm produces all the $\frac{L}{2}$
ground states when $h$ varies.

There are $n=(\frac{L}{2}-1)\frac{L}{2}$ pair of bonds at even distance, and
the sort algorithm has a complexity $n\ln n$. However in practice one is
interested only in the critical region, which means that only the pair
of bonds $(J_i,J_j)$ with $J_i+J_j$ small enough have to be considered.
This accelerates greatly the algorithm.

\section{Average correlation function in the classical limit}
\label{app:corr_func}

Let us consider a large finite chain, $L \gg 1$ in the vicinity of the random classical multicritical point $h-2 \ll 1$. Let us assume that we are in the transition regime, thus the random samples are either fully AFM ordered, thus $h< J_{i_1}+J_{i_2}$ and being a fraction $p_1$, or contain just two AFM domains $h \ge J_{i_1}+J_{i_2}$ being a fraction $p_2$, and we omit those, which contain more AFM domains, thus $p_1 + p_2 =1$. In the AFM ordered samples the average correlation function is 1. In samples with two domains for a given sample averaged correlation function is $(1-4l_1/L)$, where $l_1 \le L/2$ is the size of the smaller AFM domain: $l_1=min(|i_1-i_2|,L-|i_1-i_2|)$. The position of the smallest couplings, ${i_1}$ and ${i_2}$ are random, therefore the distribution of the domain sizes  $1\le l_1 \le L/2-1$ is uniform. Consequently for samples with two AFM domains the ensemble averaged correlation function is $0$. Then the total average of the correlation function over fully AFM ordered and two domain samples is given by $\overline{C(L/2-1,h)}=p_1=1-p_2$.

The fraction of samples with two AFM domains can be calculated through extreme-value statistics\cite{galambos,ev}. For this we should note, that a sample with two domains appear, if $h-2=(J_{i_1}-1)+(J_{i_2}-1)$, where $\epsilon_1=J_{i_1}-1$ as well as $\epsilon_2=J_{i_2}-1$ are the smallest values out of $L/2$ ones ($i_1$ and $i_2$ being of different parity) having a parent distribution, which is uniform in $[0,1]$.
According to extreme-value statistics\cite{ev}, the asymptotic form of the distribution of $\epsilon_1$, and that of $\epsilon_2$
depends on the asymptotic behavior of the parent distribution for small argument. If it is in the form: $P(\epsilon) \sim
\epsilon^{\omega}$, the scaling combination reads as $u_1=u_0 (L/2)^z \epsilon_1$, with $1/z=1+\omega$ and $u_0$ is a constant. For the uniform distribution we have $\omega=0$, thus $z=1$. The distribution of $u_1$ is given by the Fr\'echet distribution:
\be
P(u_1)=\frac{1}{z}u_1^{1/z-1}\exp\left[-u_1^{1/z}\right]\;,
\label{frechet}
\ee
and similarly for $u_2$. Then for the longitudinal field the appropriate scaling variable is $u=u_0 (L/2)^z (h-2)=u_1+u_2$ and its distribution is given as the convolution of $P(u_1)$ and $P(u_2)$, which for $z=1$ is given by:
\be
P(u)=u\exp[-u]\;.
\label{frechet2}
\ee
The fraction of samples with two domains is given by the accumulated distribution:
\be
p_2(u)=\int_0^u P(u) {\rm d}u=1-(1+u)\exp(-u)\;,
\label{p_2}
\ee
from which the result in Eq.(\ref{C_appox}) follows.

\section{Quantum corrections to the classical limit at the RCMP point}
\label{sec:Q_corrections}

Let us consider the $h>2$ part of the phase-diagram for large, but finite value of $L$, at such a point, where in the ground state of the classical model there is exactly one reversed domain, the domain boundaries being at $i=a=2\alpha+1$ and $i=L$. The couplings at the boundaries $J_a$ and $J_L$ are the smallest at the odd and even positions, respectively, and $J_a+J_L-2h<0$. The energy of the classical ground state is:
\be
E_0=E_{\rm AF}+2(J_a+J_L-2h)\;
\label{Delta_E}
\ee
where the classical AFM state has the energy: $E_{\rm AF}=-\sum_{i=1}^L J_i$.

Now let us switch on the transverse fields, and for simplicity let us consider a position independent strength: $\Gamma \ll 1$. The first non-vanishing correction to the AFM state is given by:
\be
\epsilon^{\rm AF}_2=-\sum_{j=1}^{L/2}\left[\frac{\Gamma^2}{2(J_{2j-1}+J_{2j}-h)}+\frac{\Gamma^2}{2(J_{2j}+J_{2j+1}+h)}\right]\;.
\label{corr_AFM}
\ee
The same type of corrections to the classical ground state are:
\beqn
\epsilon^{\rm 0}_2&=&-\sum_{j=1}^{\alpha-1}\left[\frac{\Gamma^2}{2(J_{2j-1}+J_{2j}-h)}+\frac{\Gamma^2}{2(J_{2j}+J_{2j+1}+h)}\right]\cr
&-&\frac{\Gamma^2}{2(J_{a-2}+J_{a-1}-h)}\cr
&-&\frac{\Gamma^2}{2(J_{a-1}-J_{a}+h)}-\frac{\Gamma^2}{2(-J_{a}+J_{a+1}+h)}\cr
&-&\sum_{j=\alpha+1}^{L/2-2}\left[\frac{\Gamma^2}{2(J_{2j}+J_{2j+1}-h)}+\frac{\Gamma^2}{2(J_{2j+1}+J_{2j+2}+h)}\right]\cr
&-&\frac{\Gamma^2}{2(J_{L-2}+J_{L-1}-h)}\cr
&-&\frac{\Gamma^2}{2(J_{L-1}-J_{L}+h)}-\frac{\Gamma^2}{2(-J_{L}+J_{1}+h)}\;.
\label{corr_CL}
\eeqn
Large contributions to the sums in Eqs.(\ref{corr_AFM}) and (\ref{corr_CL}) are due to such terms, in which $h$ in the nominator has a minus sign. In Eq.() there are $L/2$ such large terms, while in Eq.() there are just $L/2-1$. Consequently in average $\epsilon^{\rm 0}_2-\epsilon^{\rm AF}_2=\Gamma^2 C(\{J_i\},h)>0$ and with increasing $\Gamma$ the quantum correction is more and more favourable for the AFM ordered state, which at a given critical value can overcome the difference in the classical energy terms in Eq.(\ref{Delta_E}). This fact is in agreement with reentrance observed numerically in Fig.\ref{fig:dmrg_phase}.

\begin{acknowledgments}
This work was supported by the National Research Fund under Grants No. K128989, No. K115959 and No. KKP-126749. J-C. AdA extends thanks to the "Theoretical Physics Workshop" and F.I. to the Saarland University and the Institut N\'eel for supporting their visits to Budapest, Saarbr\"ucken and Grenoble, respectively.

\end{acknowledgments}


\begin{thebibliography}{99}
\bibitem{omega}
If the distribution of the small random couplings is in the form: $\pi_1(J) \sim (J-J_0)^{\omega},~(J-J_0) \ll 1$, then the appropriate scaling combination is $u=u_0 L (h-2)^{1+\omega}$. For the box-distribution in Eq.(\ref{eq:J_distrib}) $\omega=0$.
\end{thebibliography}
\end{document}